\documentclass{article}
\usepackage{arxiv}
\usepackage{graphicx}
\usepackage{mathtools}
\usepackage[utf8]{inputenc} 
\usepackage[T1]{fontenc}    
\usepackage{hyperref}       
\usepackage{url}            
\usepackage{booktabs}       
\usepackage{amsfonts}       
\usepackage{nicefrac}       
\usepackage{microtype}      
\usepackage{lipsum}

\title{Singer Identification Using Deep Timbre Feature Learning with KNN-Net}

\author{
  Xulong Zhang \\
  School of Computer Science and Technology\\
  Fudan University\\
  Shanghai, 200438 China\\
  \texttt{xlzhang14@fudan.edu.cn} \\
   \AND
  Jiale Qian \\
   School of Computer Science and Technology\\
  Fudan University\\
  Shanghai, 200438 China\\
  \texttt{jlqian20@fudan.edu.cn} \\
   \And
 Yi Yu \\
  Digital Content and Media Sciences Research Division\\ National Institute of Informatics\\
  Tokyo, 101-8430 Japan \\
  \texttt{yiyu@nii.ac.jp} \\  
    \AND
   Yifu Sun \\
   School of Computer Science and Technology\\
   Fudan University\\
   Shanghai, 200438 China\\
   \texttt{yfsun20@fudan.edu.cn} \\
   \And
   Wei Li \thanks{This work was supported by National Key R\&D Program of China(2019YFC1711800), NSFC(61671156). Corresponding author: Wei Li, weili-fudan@fudan.edu.cn}\\
   $^1$ School of Computer Science and Technology \\
   $^2$ Shanghai Key Laboratory of Intelligent Information Processing\\
   Fudan University\\
   Shanghai, 200438 China\\
   \texttt{weili-fudan@fudan.edu.cn} \\
}

\begin{document}
\maketitle

\begin{abstract}
	In this paper, we study the issue of automatic singer identification (SID) in popular music recordings, which aims to recognize who sang a given piece of song. The main challenge for this investigation lies in the fact that a singer's singing voice changes and intertwines with the signal of background accompaniment in time domain. To handle this challenge, we propose the KNN-Net for SID, which is a deep neural network model with the goal of learning local timbre feature representation from the mixture of singer voice and background music. Unlike other deep neural networks using the softmax layer as the output layer, we instead utilize the KNN as a more interpretable layer to output target singer labels. Moreover, attention mechanism is first introduced to highlight crucial timbre features for SID. Experiments on the existing artist20 dataset show that the proposed approach outperforms the state-of-the-art method by 4\%. We also create singer32 and singer60 datasets consisting of Chinese pop music to evaluate the reliability of the proposed method. The more extensive experiments additionally indicate that our proposed model achieves a significant performance improvement compared to the state-of-the-art methods.
\end{abstract}


\section{Introduction}

Automatic singer identification (SID) is considered as an increasingly important research topic in the field of audio signal processing, with the aim of recognizing who sang a given piece of song. It has many potential applications in AI music. For example, an optimal SID system can be used to recommend similar singers to users based on vocal characteristics of each singer. It is also very helpful to reduce human efforts to manage unlabeled or insufficient songs and check huge amount of suspect songs rapidly. Generally, a singer’s voice is based on the artistry of the work tends to arbitrarily change in time domain, which is significantly different from the normal audio signal. What makes the task more challenging is that the singing voice is inevitably intertwined with the background accompaniment \cite{kooshan2019singer, shen2019deep}.

Research focuses of existing SID approaches can be grouped into three categories: i) simply considering as an issue of the speaker identification (SPID) \cite{eghbal2015vectors}, ii) directly identifying singers ignoring the influence of background music for the singer voice \cite{loni2019timbre, nasrullah2019music}, and iii) distinguishing singers by voice characteristics after removing the interference of background music \cite{sharma2019importance}. Hamid et al. \cite{eghbal2015vectors} used i-vector that is first introduced in SPID task, which aims to extract song-level descriptors built from frame-level timbre features. Their experimental results achieved a F1-score of 0.846 on artist20 \cite{ellis2007classifying}, a public dataset for the SID task. Loni et al., in \cite{loni2019timbre} used the combination of timbre and vibrato features with different attributes to describe the vocal characteristics of the singer, which achieved an accuracy of 0.805 on a cappella database of 23 singers. Nasrullah et al. \cite{nasrullah2019music} proposed another supervised method using the CRNN architecture that achieved an accuracy of 0.935 on the artist20. Some new methods are proposed using singing voice separation \cite{chandna2017monoaural, jansson2017singing} as pre-processing. Sharma et al. \cite{sharma2019importance} extract the singing vocals from polyphonic songs using Wave-U-Net based approach to overcome the interference of background accompaniment, which outperforms the baseline without audio source separation by a large margin. Our method belongs to the second category.

\begin{figure*}[htb]
	\centering
	\includegraphics[width=0.9\textwidth]{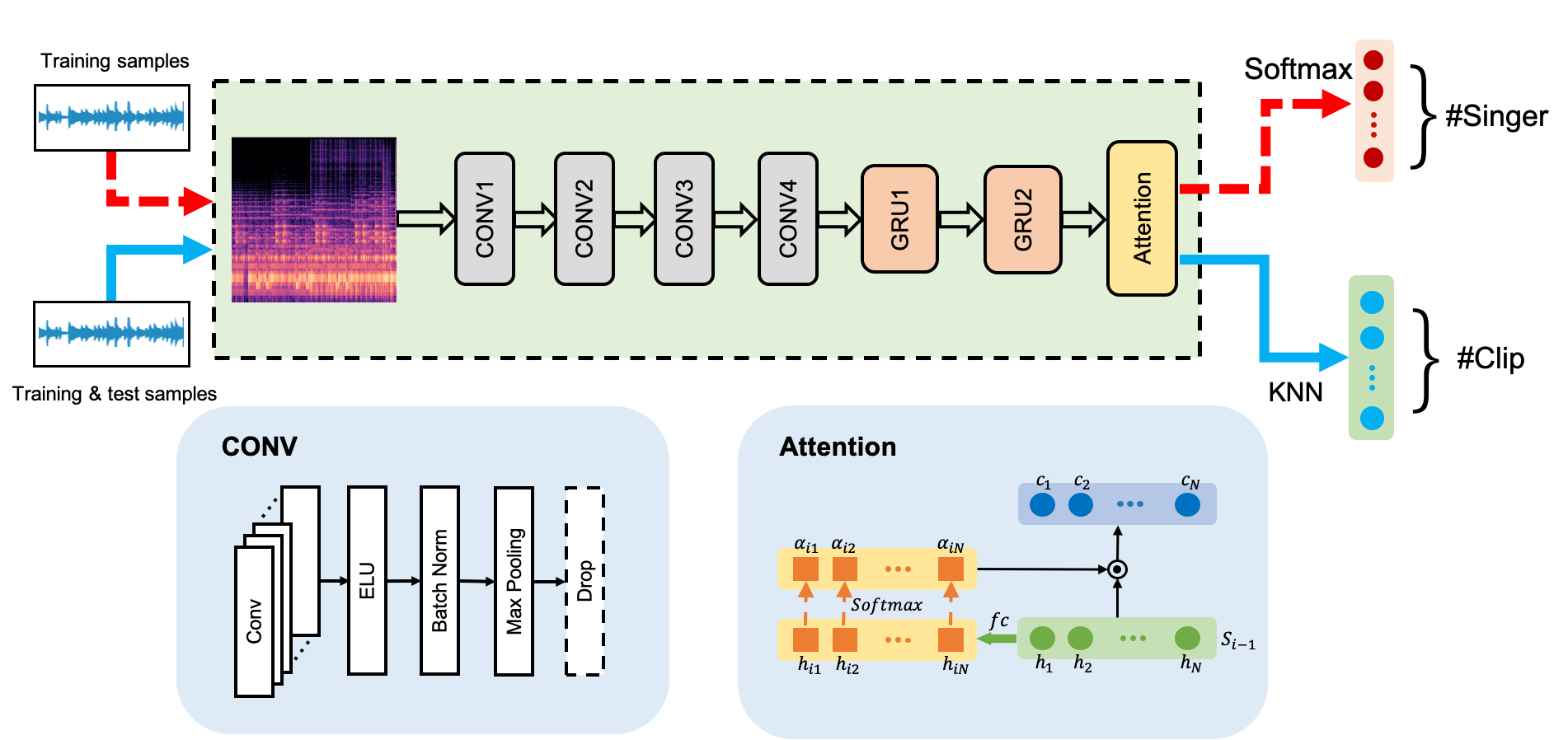}
	\caption{The overview of the proposed deep network architecture for singer identification.}
	\label{fig:1}
	
\end{figure*}

Generally, softmax layer as the output layer of DNN is the most common choice for classification tasks such as SID \cite{nasrullah2019music, alvarez2020singer, hsieh2020addressing}, but in this manner it is quite improper that all the singers are taken into account as candidates in each decision. Inspired by KNN \cite{zhang2016introduction}, a traditional machine learning method, we confine each decision to a relatively appropriate range, i.e, a fixed number of target singers or song clips. The KNN in our proposed model functions as the final layer where neighbor distances can be computed in parallel by GPU \cite{barrientos2017gpu}. As the distance matrix \cite{zhang2006svm} of KNN algorithm, timbre space is introduced to represent vocal features of each singer. Additionally, the current SID dataset is limited, which has hindered the development of large-scale SID study \cite{kooshan2019singer, nasrullah2019music, amarasinghe2016supervised}. We create our datasets named singer32 and singer60 that contain more singer labels than artist20 and evaluate our model on these datasets.

\section{Proposed Method}

The KNN-Net for SID is introduced in this section, which is a deep neural network architecture with the aim of timbre feature learning and interpretably classifying. This deep model can be extended to other audio classification tasks.

\subsection{The architecture of KNN-Net}

As the red and blue arrows presented in Fig. \ref{fig:1}, two stages are required in the proposed method. The model takes the time-frequency representation corresponding to each clip as the input. This is fed into four CNN layers to extract local timbre features, which is followed by the GRU module to extract time-domain features. After CRNN layers, an attention layer is employed to strengthen important feature representations \cite{won2019toward}.

In the first stage, the softmax layer is used to predict the probability of each singer on training dataset, which aims to train the feature extraction block. While in the next stage, softmax layer is replaced by a linear dense layer which is also defined as the KNN layer, and the weights of the attention and CRNN layers are fixed. The KNN layer following attention layer then returns the high-dimensional cosine distance vectors as the output. Subsequently, top K song clips are chosen according to these cosine distance vectors, where k represents the threshold value of the KNN algorithm. Target singer labels are finally estimated by voting the most frequent singer among the top K song clips.

The workflow of the proposed deep network architecture for SID is summarized as follows:
i) Learning timbre space based on attention-CRNN model.
ii) Using the model learned in i) to output the reference matrix of training samples, which represents timbre features of each singer.
iii) Using the reference matrix as the input weight to construct a dense layer without bias which is also called a KNN layer.
iv) Constructing KNN-Net by integrating learned feature extraction component and the KNN layers without retraining.
v) Predicting target singer labels with KNN-Net, which aims to search the most similar singers in the timbre space.

\subsection{Attention mechanism for timbre space learning}

The differences between singers lie in their vocal characteristics and frequency bands of spectral features \cite{zhang2016end}. Typically, voice information of male singers mainly is in the low and medium frequency bands while that of female singers tends to be in higher frequency bands \cite{orman2001frequency}, which indicates the importance of the strategy to give different frequency bands with different attention weights. For example, the network should learn to focus on low and medium frequency bands when handling the songs performed by male singers and the higher frequency bands for female singers. Consequently, we introduce the attention layer to explore the contribution of each frequency band to the whole task and highlight key parts \cite{yuan2019fusionatt}.

The structure of attention layer is presented in Fig. \ref{fig:1}. Assuming that a sequence of N units is returned from the GRU module, the output of the attention layer could be represented as
\begin{eqnarray}
c_i = \sum_{j=1}^{N}\alpha_{ij}h_j \label{eq:6}
\end{eqnarray}

\noindent where $c_i$ represents the hidden vector obtained from weighted summation of $S_{i-1}=\{h_1,...,h_N\}$, which represents the hidden vector of the output from the last GRU layer. The attention weight of each feature vector can be obtained by equation (\ref{eq:7}) and equation (\ref{eq:8}):
\begin{eqnarray}
\alpha_{ij} = \frac{exp(h_{ij})}{\sum_{k=1}^{N}exp(h_{ik})} \label{eq:7}
\end{eqnarray}

\begin{eqnarray}
h_{ij} = fc(S_{i-1}, h_j) \label{eq:8}
\end{eqnarray}

\noindent where $h_{ij}$ represents the score of dependence relationship between $S_{i-1}$ and $h_j$, which is obtained by the learned mapping function $fc$. After normalization using the softmax function shown in equation (\ref{eq:7}), the final output of attention layer is obtained as the weighted sum.

The attention block is independent from the mixed model of CNN and GRU, which captures the important features extracted by CRNN and strengthens their relationship.

\begin{figure}[htb]
	
	\begin{minipage}[b]{1.0\linewidth}
		\centering
		\centerline{\includegraphics[width=7cm]{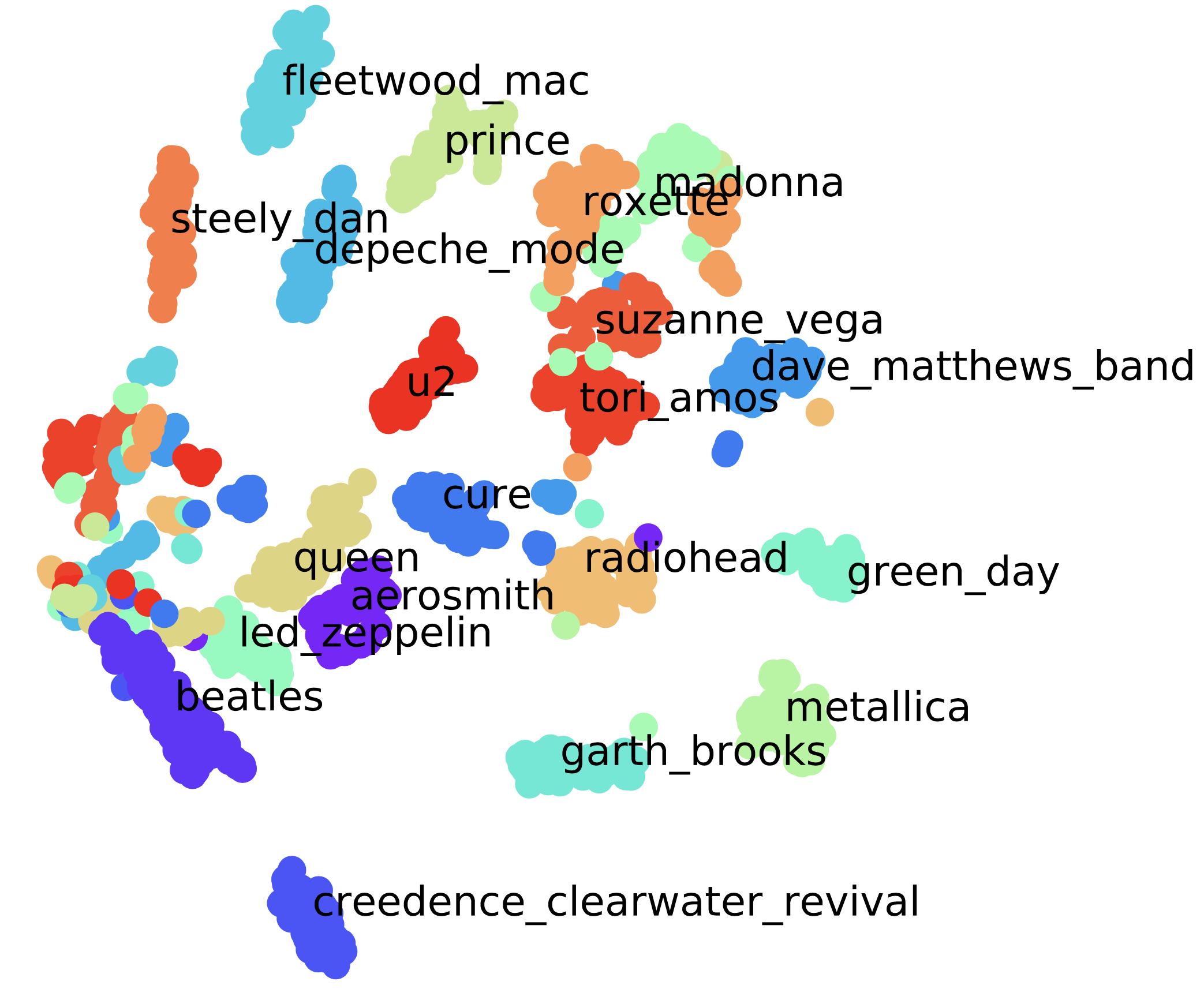}}
		\caption{Visualization of the timbre features of each singer using t-SNE. Singers with similar timbre features and music genre are tend to be close, e.g., the lead singers of Led Zepplin and Aerosmith.}
		\label{fig:2}
	\end{minipage}
\end{figure}

\subsection{Distance computation in the KNN layer}

Traditional KNN method is integrated into the deep learning model, and the key idea is to design a network structure related to distance measurement. The following is the detailed derivation of the distance computation in the KNN layer.

Let $q$ be the feature vector of test samples extracted by the attention CRNN network, $\mathbf{W}$ be the reference matrix obtained from training samples and $w$ represents one specific column of $\mathbf{W}$. According to the cosine distance measure \cite{chomboon2015empirical}, the distance between $q$ and $w$ can be easily computed as
\begin{eqnarray}
cos(q, w) = \frac{q \cdot w}{\lvert q \rvert \cdot \lvert w \rvert}   \label{eq:1}
\end{eqnarray}

For each given test sample, $\lvert q \rvert$ is regarded as a constant value. After normalization operation ($\lvert w \rvert=1$), the cosine distance in equation (\ref{eq:1}) can be converted as 
\begin{eqnarray}
cos(q, \mathbf{W}) = q \cdot \mathbf{W}   \label{eq:3}
\end{eqnarray}

In the neural network, the reference matrix $\mathbf{W}$ is passed into the dense layer as
\begin{eqnarray}
\mathbf{A} = g(q \cdot \mathbf{W} + b)   \label{eq:4}
\end{eqnarray}

\noindent where $\mathbf{A}$ represents the result matrix and $g$ represents the activation function, respectively. When $g$ is a linear function and b is 0, $\mathbf{A}$ can be computed as
\begin{eqnarray}
\mathbf{A} = q \cdot \mathbf{W}   \label{eq:5}
\end{eqnarray}

Equation (\ref{eq:5}) referring to the dense layer is therefore equal to equation (\ref{eq:3}), which indicates that the dense layer can be interpreted as the KNN method.

\begin{table*}
	\centering
	\caption{Evaluation on artist20, singer32 and singer60: The results of the attention-CRNN-KNN (KNN-Net) achieve the highest accuracy compared to the baseline, and the accuracy on artist20 is significantly higher than other two datasets.} \label{table:1}
	\begin{tabular}{rcccccccccccc}
		\hline
		Measurement & \multicolumn{3}{c}{Accuracy} & \multicolumn{3}{c}{Precision} & \multicolumn{3}{c}{Recall} & \multicolumn{3}{c}{F1} \\ \hline
		Dataset            & A20  & S32  & S60  & A20  & S32  & S60  & A20  & S32  & S60  & A20  & S32  & S60  \\ \hline
		i-vector \cite{eghbal2015vectors}              & 0.85 & - & - & 0.86 & - & - & 0.86 & - & - & 0.85 & - & - \\
		SVS-SID \cite{sharma2019importance}              & 0.90 & - & - & - & - & - & - & - & - & - & - & - \\
		CRNN \cite{nasrullah2019music}               & 0.94 & 0.69 & 0.57 & 0.93 & 0.72 & 0.57 & 0.93 & 0.69 & 0.56 & 0.93 & 0.68 & 0.51 \\
		Attention-CRNN     & 0.95 & 0.72 & 0.63 & 0.96 & 0.75 & 0.62 & 0.95 & 0.69 & 0.62 & 0.95 & 0.69 & 0.58 \\
		Attention-CRNN-KNN & 0.99 & 0.74 & -    & 0.99 & 0.76 & -    & 0.99 & 0.70 & -    & 0.99 & 0.71 & -    \\ \hline
	\end{tabular}
\end{table*}

\section{Experiments and Results}

Our experiments are designed to investigate whether our proposed KNN-Net model achieves a performance comparable with previous methods, and to measure the effect of the introduced attention mechanism.

\subsection{Dataset}

We use the artist20 (A20) dataset that contains 20 singer labels and 1413 songs. Each singer has 6 albums in which 4 albums are used in the training set and the remaining 2 albums are used in the validation and test set. We extend artist20 into singer32\footnote{\url{https://zenodo.org/record/3822288}} (S32) and singer60\footnote{\url{https://zenodo.org/record/3823866}} (S60). Singer32 contains 32 Chinese pop singers, and each singer has 70 music recordings which are extracted from their music videos. Similarly, singer60 consists of 60 singers, including 30 male singers and 30 female singers, and each singer has 70 songs. The collections of singer32 and singer60 are divided into training, validation and test sets in the ratio of 8:1:1.

\subsection{Evaluation Metrics}

In order to quantitatively evaluate the classification performance of the proposed model, the accuracy, precision, recall and F1 score are taken as the performance metrics.

\subsection{Experimental Setup}

As convolutional recurrent neural network (CRNN) proposed in \cite{nasrullah2019music} performed best results in the existing state-of-art works of SID, we took it as our comparison method. In addition, we also add attention mechanism to CRNN and treat it as another important comparison method. Some results obtained by i-vector \cite{eghbal2015vectors} and SVS-SID \cite{sharma2019importance} on artist20 dataset are taken as comparisons. 

In our method, target singer labels are first estimated based on the frame-block size (takes up 1 second that consists of 32 frames). The recognition result of an entire song can be voted from all blocks. The detailed settings of network layers and parameters of CRNN and attention-CRNN are currently available at gitlab\footnote{\url{https://gitlab.com/exp\_codes/tut-attention}}. And the value of K in the proposed KNN-Net method is set as 11 based on some initial experiments.

\subsection{Results and Discussion}

\begin{figure}[htb]
	
	\begin{minipage}[b]{1.0\linewidth}
		
		\centerline{\includegraphics[width=8.5cm]{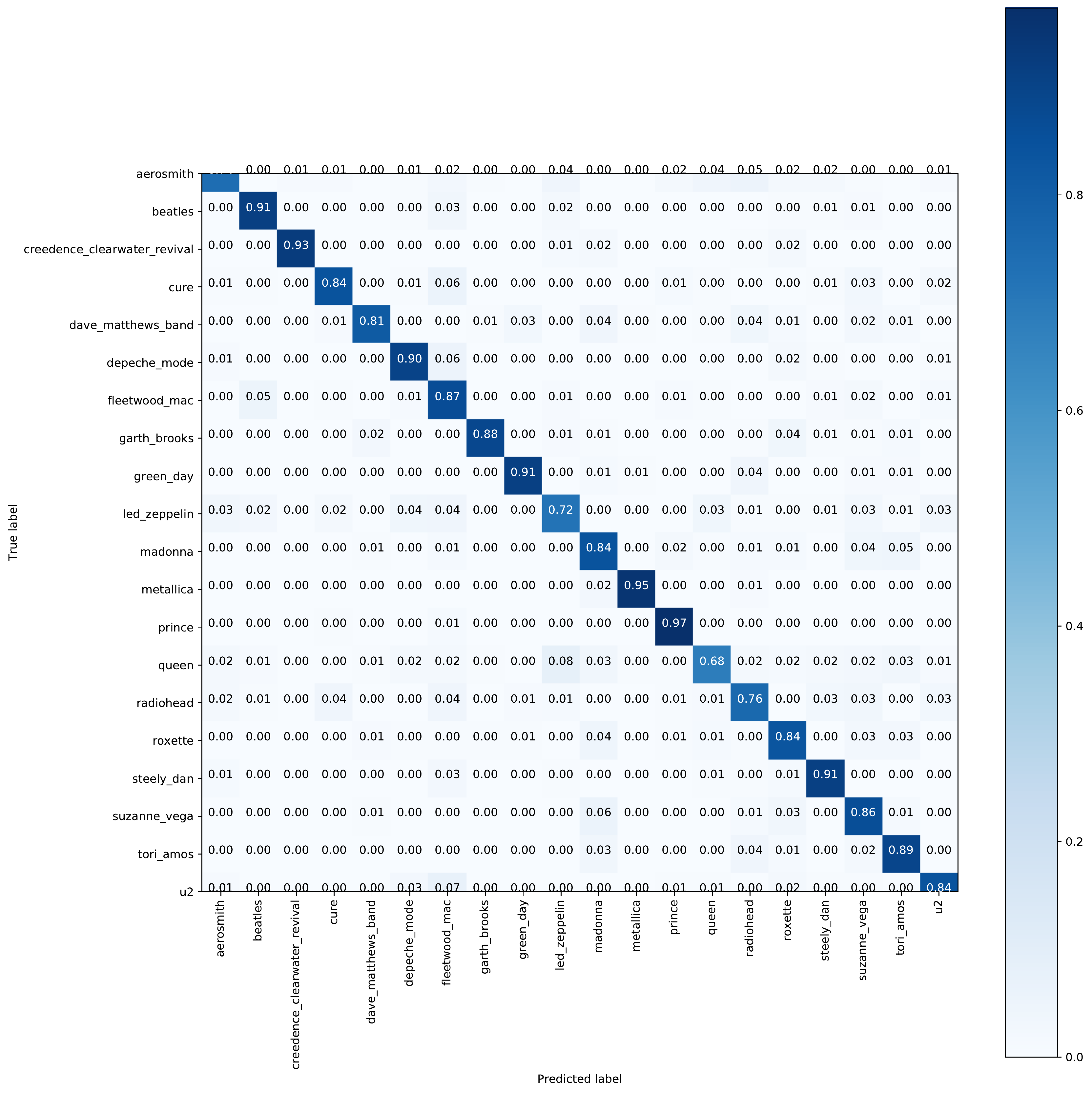}}
		\caption{The confusion matrix of the recognition results based on frame-block level evaluated on artist20.}
		\label{fig:4}
	\end{minipage}
	
\end{figure}

According to the experimental results on the public dataset artist20, we present the confusion matrix of the recognition results of the proposed method based on frame-block level as shown in Fig. \ref{fig:4}, wherein the recognition accuracy has an average of 85$\%$.

Table \ref{table:1} shows the singer identification results for our proposed model and comparison methods. It is observed that our KNN-based deep architecture achieves the best performance, improving the performance  significantly compared with the baseline model. Specifically, for artist20, a little improvement in the recognition accuracy is achieved when introducing the attention mechanism to CRNN, while a 5\% increase is achieved when using our attention-CRNN-KNN model. Obviously, the experimental results have higher accuracy on the dataset artist20 than on the other two datasets. For artist20, the relatively small capacity of the dataset is one of the reasons. In addition, the dataset also contains music pieces of various genres that provide additional information to the neural network, which may improve the recognition accuracy to a certain extent. Comparatively, singer32 and singer60 include only Chinese pop music and do not contain any genre-related information. Unfortunately, due to the problem of large dimension of the reference matrix, we can not give the experimental results on singer60.

\section{Conclusion}

We have presented a novel KNN-based deep architecture that learns the timbre feature extraction network to recognize singers by KNN approach, which can be regarded as a query searching task based on timbre feature space. Moreover, a large-scale reference matrix is computed in parallel efficiently by the cosine distance computation with the dense layer. We evaluated our method on three datasets for singer identification and confirmed that it outperforms the state-of-the-art methods.

Considering the problem of large dimensional reference matrix that cannot be stored in the memory when enlarging the dataset, the future work is to implement a centralization method on the reference matrix. We would like to utilize a clustering method to reduce the dimension of the reference matrix so that the requirements of large-scale singer identification could be satisfied.


\bibliographystyle{unsrt}
\bibliography{references}

\begin{thebibliography}{10}

\bibitem{kooshan2019singer}
Seyed Kooshan, Hashemi Fard, and Rahil~Mahdian Toroghi.
\newblock Singer identification by vocal parts detection and singer
  classification using lstm neural networks.
\newblock In {\em Proceedings of the 4th International Conference on Pattern
  Recognition and Image Analysis}, pages 246--250. IEEE, 2019.

\bibitem{shen2019deep}
Zebang Shen, Binbin Yong, Gaofeng Zhang, Rui Zhou, and Qingguo Zhou.
\newblock A deep learning method for chinese singer identification.
\newblock {\em Tsinghua Science and Technology}, 24(4):371--378, 2019.

\bibitem{eghbal2015vectors}
Hamid Eghbal-Zadeh, Bernhard Lehner, Markus Schedl, and Gerhard Widmer.
\newblock I-vectors for timbre-based music similarity and music artist
  classification.
\newblock In {\em Proceedings of the 16th International Society for Music
  Information Retrieval Conference}, pages 554--560, 2015.

\bibitem{loni2019timbre}
Deepali~Y Loni and Shaila Subbaraman.
\newblock Timbre-vibrato model for singer identification.
\newblock {\em Information and Communication Technology for Intelligent
  Systems}, pages 279--292, 2019.

\bibitem{nasrullah2019music}
Zain Nasrullah and Yue Zhao.
\newblock Music artist classification with convolutional recurrent neural
  networks.
\newblock In {\em Proceedings of the 2019 International Joint Conference on
  Neural Networks}, pages 1--8. IEEE, 2019.

\bibitem{sharma2019importance}
Bidisha Sharma, Rohan~Kumar Das, and Haizhou Li.
\newblock On the importance of audio-source separation for singer
  identification in polyphonic music.
\newblock In {\em Proceedings of the 20st Annual Conference of the
  International Speech Communication Association}, pages 2020--2024, 2019.

\bibitem{ellis2007classifying}
Daniel~PW Ellis.
\newblock Classifying music audio with timbral and chroma features.
\newblock In {\em Proceedings of the 8th International Conference on Music
  Information Retrieval}, 2007.

\bibitem{chandna2017monoaural}
Pritish Chandna, Marius Miron, Jordi Janer, and Emilia G{\'o}mez.
\newblock Monoaural audio source separation using deep convolutional neural
  networks.
\newblock In {\em Proceedings of the International conference on latent
  variable analysis and signal separation}, pages 258--266. Springer, 2017.

\bibitem{jansson2017singing}
Andreas Jansson, Eric Humphrey, Nicola Montecchio, Rachel Bittner, Aparna
  Kumar, and Tillman Weyde.
\newblock Singing voice separation with deep u-net convolutional networks.
\newblock In {\em Proceedings of the 18th International Society for Music
  Information Retrieval Conference}, 2017.

\bibitem{alvarez2020singer}
Aitor~Arronte Alvarez and Francisco Gomez-Martin.
\newblock Singer identification using convolutional acoustic motif embeddings.
\newblock {\em arXiv preprint arXiv:2008.00198}, 2020.

\bibitem{hsieh2020addressing}
Tsung-Han Hsieh, Kai-Hsiang Cheng, Zhe-Cheng Fan, Yu-Ching Yang, and Yi-Hsuan
  Yang.
\newblock Addressing the confounds of accompaniments in singer identification.
\newblock In {\em Proceedings of the 2020 IEEE International Conference on
  Acoustics, Speech and Signal Processing}, pages 1--5, 2020.

\bibitem{zhang2016introduction}
Zhongheng Zhang.
\newblock Introduction to machine learning: k-nearest neighbors.
\newblock {\em Annals of translational medicine}, 4(11), 2016.

\bibitem{barrientos2017gpu}
Ricardo~J Barrientos, Fabricio Millaguir, Jos{\'e}~L S{\'a}nchez, and Enrique
  Arias.
\newblock Gpu-based exhaustive algorithms processing knn queries.
\newblock {\em The Journal of Supercomputing}, 73(10):4611--4634, 2017.

\bibitem{zhang2006svm}
Hao Zhang, Alexander~C Berg, Michael Maire, and Jitendra Malik.
\newblock Svm-knn: Discriminative nearest neighbor classification for visual
  category recognition.
\newblock In {\em Proceedings of the IEEE Computer Society Conference on
  Computer Vision and Pattern Recognition}, volume~2, pages 2126--2136. IEEE,
  2006.

\bibitem{amarasinghe2016supervised}
Rajitha Amarasinghe and Lakshman Jayaratne.
\newblock Supervised learning approach for singer identification in sri lankan
  music.
\newblock {\em European Journal of Computer Science and Information
  Technology}, 4(6):1--14, 2016.

\bibitem{won2019toward}
Minz Won, Sanghyuk Chun, and Xavier Serra.
\newblock Toward interpretable music tagging with self-attention.
\newblock {\em arXiv preprint arXiv:1906.04972}, 2019.

\bibitem{zhang2016end}
Shi-Xiong Zhang, Zhuo Chen, Yong Zhao, Jinyu Li, and Yifan Gong.
\newblock End-to-end attention based text-dependent speaker verification.
\newblock In {\em 2016 IEEE Spoken Language Technology Workshop}, pages
  171--178. IEEE, 2016.

\bibitem{orman2001frequency}
{\"O}zg{\"u}r~Devrim Orman and Levent~M Arslan.
\newblock Frequency analysis of speaker identification.
\newblock In {\em A Speaker Odyssey - The Speaker Recognition Workshop}, pages
  219--222, 2001.

\bibitem{yuan2019fusionatt}
Ye~Yuan and Kebin Jia.
\newblock Fusionatt: deep fusional attention networks for multi-channel
  biomedical signals.
\newblock {\em Sensors}, 19(11):2429, 2019.

\bibitem{chomboon2015empirical}
Kittipong Chomboon, Pasapitch Chujai, Pongsakorn Teerarassamee, Kittisak
  Kerdprasop, and Nittaya Kerdprasop.
\newblock An empirical study of distance metrics for k-nearest neighbor
  algorithm.
\newblock In {\em Proceedings of the 3rd international conference on industrial
  application engineering}, pages 280--285, 2015.

\end{thebibliography}

\end{document}